\documentclass[11pt]{article}
\usepackage{amsmath}
\usepackage{bm}
\usepackage{graphicx,epsfig}
\usepackage{geometry}
\makeatletter
\renewenvironment{thebibliography}[1]
{\section*{\refname\@mkboth{\refname}{\refname}}%
  \list{\@biblabel{\@arabic\c@enumiv}}%
       {\settowidth\labelwidth{\@biblabel{#1}}%
        \leftmargin\labelwidth
        \advance\leftmargin\labelsep
 \setlength\baselineskip{11pt}%
        \@openbib@code
        \usecounter{enumiv}%
        \let\p@enumiv\@empty
        \renewcommand\theenumiv{\@arabic\c@enumiv}}%
  \sloppy
  \clubpenalty4000
  \@clubpenalty\clubpenalty
  \widowpenalty4000%
  \sfcode`\.\@m}
 {\def\@noitemerr
 {\@latex@warning{Empty `thebibliography' environment}}%
\endlist}
\makeatother
\begin{document}
\centerline{{\sl Genshikaku Kenkyu Suppl.} No. 000 (2012)}
\begin{center} 
\vskip 2mm
{\Large\bf
%
p-Shell Hypernuclear Spectroscopy at JLab's Hall A \\ and Future 
Prospects 
\hspace{-1mm}\footnote{Presented at the International Workshop on Strangeness 
Nuclear Physics (SNP12), August 27 - 29, 2012, \\
\hspace*{5mm} Neyagawa, Osaka, Japan.}
}\vspace{5mm}

{
F.~Garibaldi$^{a,b}$, P.~Byd\v{z}ovsk\'y$^c$, E.~Cisbani$^b$, 
F.~Cusanno$^d$, 
R.~De~Leo$^e$, S.~Frullani$^f$, D.~W.~Highinbotham$^g$, M.~Iodice$^h$, 
J.~J.~LeRose$^g$, P.~Markowitz$^l$, G.~M.~Urciuoli$^a$, \\ 
for the Hall A Collaboration
}\bigskip 

{\small
$^a$Istituto Nazionale di Fisica Nucleare, Sezione di Roma, I-00185 
Rome, Italy \\ 
$^b$Istituto Superiore di Sanit\`a, I-00166 Rome, Italy \\
$^c$Nuclear Physics Institute, 250 68 \v{R}e\v{z} near Prague, Czech 
Republic\\ 
$^d$Technische Universit\"{a}t M\"{u}nchen, Excellence Cluster Universe, 
D-85748 Garching, Germany\\ 
$^e$Istituto Nazionale di Fisica Nucleare, Sezione di Bari, I-70125 Bari, 
Italy\\
$^f$Istituto Nazionale di Fisica Nucleare, Gruppo Collegato Sanit\`a 
Sezione di Roma, I-00166 Rome, Italy\\
$^g$Thomas Jefferson National Accelerator Facility, Newport News VA 
23606, USA\\
$^h$Istituto Nazionale di Fisica Nucleare, Sezione di Roma Tre, I-00146 
Rome, Italy\\
$^l$Florida International University, Miami FL 32306, USA\\
}
\end{center}
\vspace{3mm}

\noindent
{\small \textbf{Abstract}:\quad
The first systematic study of 1p-shell and medium-heavy hypernuclei 
by electroproduction of strangeness has started at Jefferson Laboratory with 
the experiments E89-009, E94-107, E01-011, E05-115. 
The main results obtained in Hall A and future prospects 
of the investigation of hypernuclei at Jefferson Laboratory regarding 
the study of the angular dependence of electroproduction of
strangeness and the possibility of performing the spectroscopy of
$^{208}_{\Lambda}Tl$ are reported here.
}%


\section{Introduction}

The physics of hypernuclei is an important branch of contemporary 
nuclear physics. Indeed, due to the low intensities of the available 
$\Lambda$ beams, only very limited information about $\Lambda-N$ 
interaction can be extracted by means of $\Lambda-p$ low energy 
scattering, therefore the hypernuclei provide a unique laboratory for 
studying the $Y-N$ force. Many progresses have been done in the last years 
in the physics of hypernuclei and few facilities in the world are planning 
new challenging experiments in this field \cite{Botta}.\\ 
In principle, information derived from 
hypernuclear structures have important implications in the study of 
the core of neutron stars, providing constraints to the equation of 
state \cite{Schulze, Schaffner}. \\
In the case of production of hypernuclei by electromagnetic probes 
\cite{HashTam} the parameters of the in medium effective 
$\Lambda-N$ potential could be determined from the structure of 
the missing mass spectra of the $^AZ(e,e'K^+)^A_{\Lambda}(Z-1)$ 
reactions. Therefore, those experiments are complementary to the 
investigations using hadronic probes \cite{Finuda} which produce mirror 
hypernuclei, hence the Charge Symmetry Breaking (CBS) term of 
the $\Lambda-N$ interaction could be accessed. 
In addition, experiments using 
electromagnetic and hadronic probes are complementary to the 
successful studies performed by $\gamma$-ray spectroscopy 
\cite{Tamura} since some energy levels can not be determined 
by $\gamma$-ray decays.\\ 
Theoretical predictions reported here of the electroproduction of 
hypernuclei are obtained in the framework of the 
Distorted Wave Impulse Approximation 
(DWIA) \cite{Sotona} using the Saclay-Lyon (SLA) model \cite{SLA} for 
the elementary process. 

\section{Electroproduction of Strangeness at Jefferson Laboratory}
\begin{figure}[htb!]
\centering
\includegraphics[width=9.cm, angle=270]{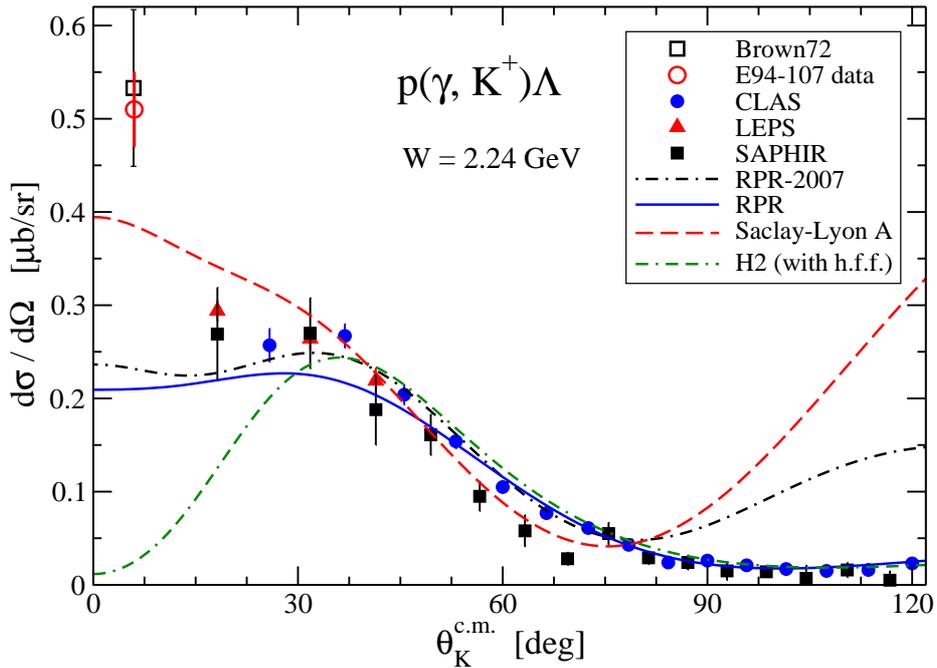}
\caption{\small \label{Fig1Label}
Cross section of photoproduction of strangeness as function of the
$K^+$ scattering angle. See the text for details.
}
\end{figure}

The beam and 
the spectrometers 
available at Jefferson Laboratory (JLab) are well suited for studying 
electroproduction of strangeness. Last decade both Hall A and Hall C 
started the studies of electroproduction spectra of 
hypernuclei with sub-MeV resolution \cite{Miyoshi, Yuan}. To some 
extent, 
the systematical study of hypernucleus spectra can provide information 
also on the elementary process of electroproduction of strangeness. This 
possibility is evident in the peculiar case of using a waterfall target 
\cite{waterfall}. 
The elementary amplitude for kaon electroproduction on the proton 
can be probed by studying the hypernucleus-production cross sections 
for various excited states in DWIA as they are sensitive 
to the specific kinematical 
region at very small kaon angles. This kinematical region is not well 
understood either from the experimental nor the theoretical side as is 
shown in Fig.~\ref{Fig1Label}.
Direct measurement of the elementary cross section is problematic and 
the lack of data do not allow thoroughly testing models for zero kaon angles. 
This results in a wide range of model predictions as seen in 
Fig.~\ref{Fig1Label}.

In Fig.~\ref{Fig1Label} predictions from the isobar and 
Regge-plus-resonance models are compared with data for 
photoproduction (CLAS~\cite{CLAS05}, LEPS~\cite{LEPS}, 
SAPHIR~\cite{SAPHIR03}), and electroproduction (E94-107~\cite{Pete} 
and Brown72~\cite{Brown}). 
For $\theta_K^{c.m.} < 30^\circ$ dynamics of the models is not well 
determined yet. In general, the models  
group into three different variants. The isobar model with hadronic form 
factors (H2~\cite{HYP03}) is strongly suppressed at zero $\theta_K$ which 
is ruled out by a comparison of the hypernucleus cross sections as predicted 
by H2 with experimental data~\cite{Iodice,Cusanno}. 
On the other hand, the isobar model without the form factors 
(SLA~\cite{SLA}) 
is consistent with the hypernucleus cross sections \cite{Iodice, Cusanno}
and also with the measurement of the elementary electroproduction cross section 
in Hall A~\cite{Pete} and the older measurement published in \cite{Brown} 
(Brown72 in Fig.~\ref{Fig1Label}). 
The hybrid models based on the Regge and isobar formalisms, the old version 
RPR-2007~\cite{Tamara} and our recent fit RPR, predict a flat angular 
dependence below $30^\circ$  which is quite consistent with the 
photoproduction 
data, see Fig.~\ref{Fig1Label}.\\
The E94-107 data reported in Fig.~\ref{Fig1Label} were measured at
JLab's Hall A using 3.777~GeV electron beam with central values of
$W=2.2~GeV$ and $Q^2=0.07~(GeV/c)^2$~\cite{Iodice} which are very similar 
to those for Brown72 data, $W=2.17~GeV$ and 
$Q^2=0.18~(GeV/c)^2$~\cite{Brown}. 
The measured cross sections for electroproduction are
higher than for photoproduction, hence it is possible that,
although at a low $Q^2$, longitudinal and interference response 
functions could strongly contribute to the cross section.

Precise measurements of the energy and angular dependence of the  
cross sections in the kaon electroproduction on the proton in the very 
small kaon-angle region, e.g. using liquid-$H_2$ or waterfall target, 
would help in determining the forward angle dynamics of the process. 
\subsection{Experimental Equipment in Hall A}
The CEBAF accelerator at JLab delivers very high intensity continuous 
electron beam to its experimental halls. 
This condition is a mandatory requirement for 
electroproduction of strangeness due to its small cross section. In 
addition, the very good energy spread and the precise determination 
of the absolute central energy of the CEBAF beam allow high-quality 
spectroscopy of hypernuclei. \\
In Hall A, the detector \cite{HallA} is based on the two-arm 
High-Resolution-Spectrometer (HRS) having a momentum resolution of 
$10^{-4}$ which is sufficient for obtaining energy resolution of few 
hundreds of keV in missing mass spectroscopy. In order to increase the 
counting rates, scattering reactions at low $Q^2$ have to be performed, 
hence the electron scattering angle has to be as small as possible and 
the kaon direction must be close to the direction of the virtual photon. 
These conditions mean that small scattering angles correspond to larger 
counting rates. Since the standard setup of HRS has a minimal angle to 
the beam axes of 
$12.5^{\circ}$, two superconducting septum magnets \cite{Guido} were 
added to the experimental setup 
in order to allow electron scattering angle and kaon angle as small as 
$6^{\circ}$, corresponding to $Q^2=0.08~(GeV/c)^2$ in the 
kinematics adopted for E94-107 \cite{Iodice}. 
At such forward angle the background of $\pi^+$ and $p$ in the $K^+$ arm is 
huge, requiring an excellent particle identification (PID). 
For this purpose the standard PID of HRS in the hadron arm was 
improved adding a RICH detector \cite{RICH}.

\subsection{Electroproduction of Hypernuclei in Hall A}
\begin{figure}[htb!]
\centering
\includegraphics[width=12cm, angle=0]{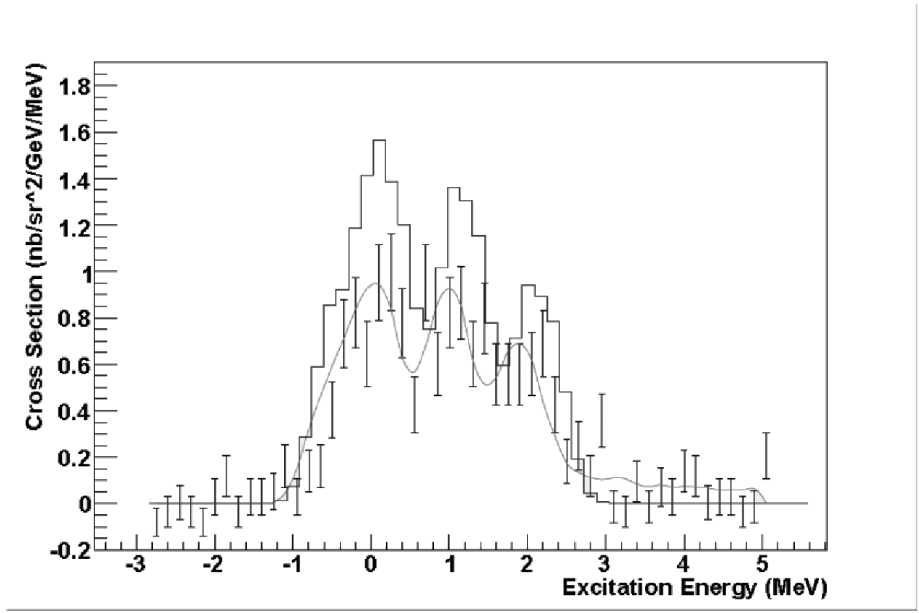}
\caption{\small \label{RadiativeCorr}
Correction of radiative effects on preliminary $^{9}_{\Lambda}Li$ 
excitation energy spectrum. The 
black points are the experimental data, the gray curve represents 
simulated data 
including radiative effects, the histogram represents simulated 
data with no radiative effects. See text for details. 
}
\end{figure}

\begin{figure}[htb!]
\centering
\includegraphics[width=9cm, angle=270, trim=0 0 0 0,clip]
{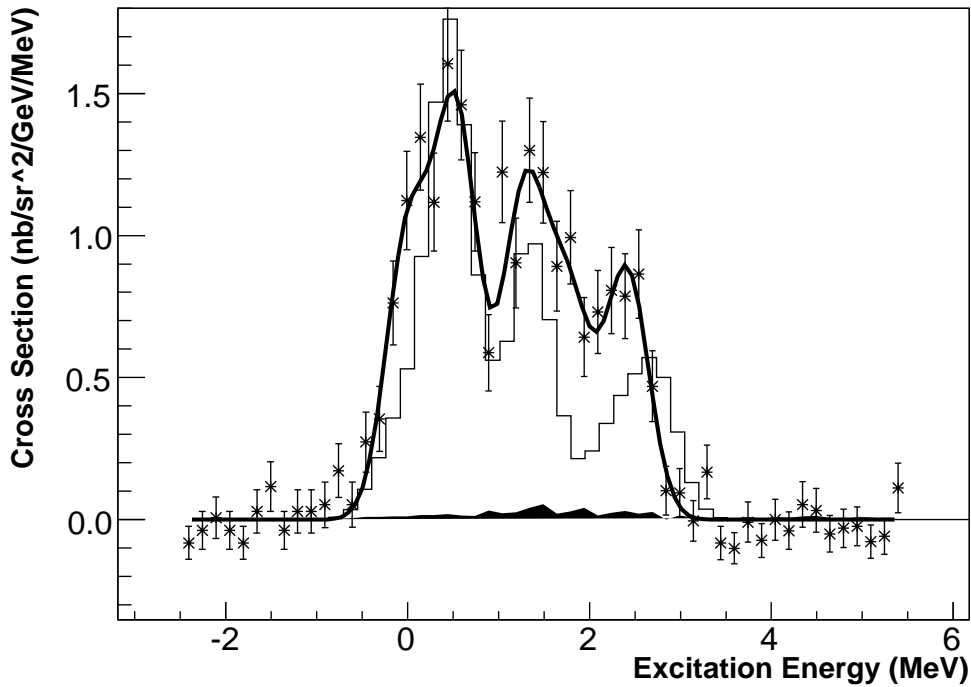}
\caption{\small \label{Fig2Label}
Five-peak gaussian fit on preliminary $^{9}_{\Lambda}Li$ excitation 
energy spectrum corrected for radiative effects. Data points 
are reported with their statistical errors, the black-filled 
histogram represents the systematical errors, the thin-black-line 
histogram represents the expectations from the theoretical model 
\cite{Sotona}, the thick black curve is a 
fit of the experimental data as explained in the text. 
}
\end{figure}

\begin{figure}[htb!]
\centering
\includegraphics[width=9cm, angle=270, 
trim = 0 0 0 0,clip]{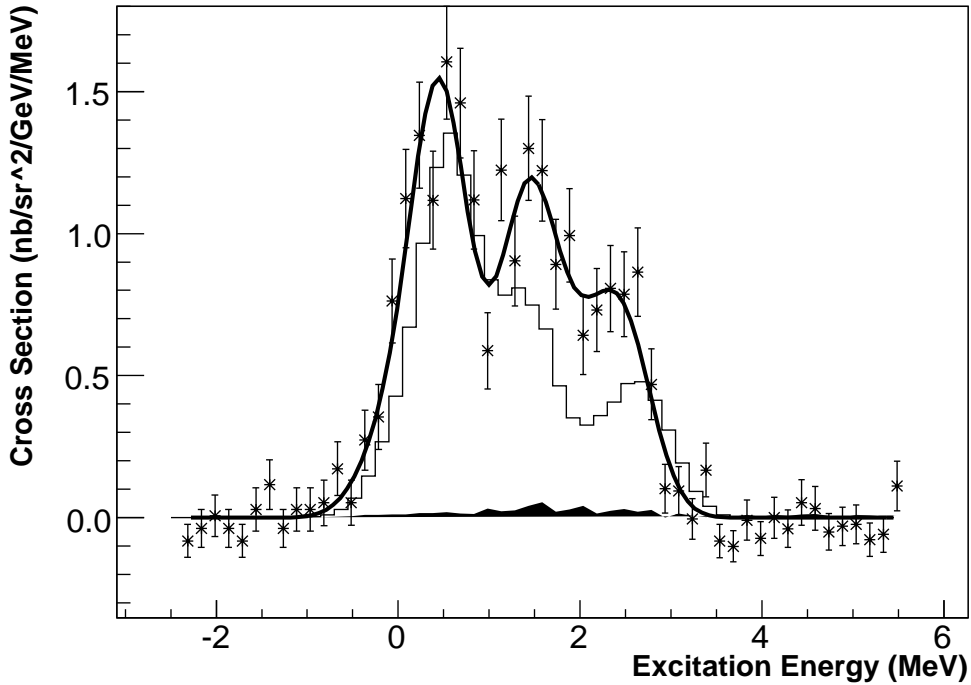}
\caption{\small \label{Fig3Label}
Same as Fig.~\ref{Fig2Label} in case of constraining the fit according 
to the theoretical model. See text for details. For the theoretical 
histogram the width of 760 keV (FWHM) was used contrary to 570 keV 
in Fig.~\ref{Fig2Label}.
}
\end{figure}

Experiment E94-107 used $^{12}C$ and $^9Be$ solid targets and a 
waterfall target in order to study respectively the 
$^{12}C(e,e'K^+)^{12}_{\Lambda}B$, the 
$^{9}Be(e,e'K^+)^{9}_{\Lambda}Li$, and the
$^{16}O(e,e'K^+)^{16}_{\Lambda}N$ reactions.
Detailed analysis of the $^{12}_{\Lambda}B$ excitation energy 
spectrum and 
$^{16}_{\Lambda}N$ binding energy spectrum are found respectively in 
\cite{Iodice} and 
\cite{Cusanno}. The analysis of $^{9}_{\Lambda}Li$ energy spectrum is 
not yet finalized: 
with respect to what reported in \cite{HypX} a fine correction for 
radiative effects has been performed,
Fig.~\ref{RadiativeCorr} shows the effect of this correction.
The Monte Carlo code SIMC \cite{SIMC} has been used for this purpose: 
once the simulated 
data fit well the experimental data, then the radiative effects 
in SIMC are turned off and the ratios between simulated data with no 
radiative effects and simulated data including radiative effects are the 
bin-by-bin correction factors for the experimental points. 
\\
Fig.~\ref{Fig2Label} shows a preliminary 
$^{9}_{\Lambda}Li$ excitation energy spectrum corrected for the 
radiative effects. 
Since the DWIA calculations predict five states, a five-peak gaussian 
fit (thick black curve) is performed on the data 
points, with the only constraint of having the same width for the five peaks. 
The resulting width for the five peaks is 570~keV (FWHM), 
then the histogram of the predicted values is obtained assuming the 
same width for the theoretical expectations (thin line).  
\\
Fig.~\ref{Fig3Label} shows the same data as in Fig.~\ref{Fig2Label} 
but a different fit is calculated. Indeed in this case the five-peak 
gaussian fit is constrained by the model: the separation and the 
relative amplitude of the individual levels composing the first and 
the third peak (doublets) are fixed according to the theoretical 
expectations. In other words, this fit corresponds to a 
three-peak fit where the internal structure of the complex peaks is 
determined by the theory.
As well as for Fig.~\ref{Fig2Label}, the 
width of the five peaks is constrained to be a single value, resulting 
here in 760~keV (FWHM).\\ 
According to this preliminary analysis, the position of the peaks and 
the amplitude of the first doublet are in 
good agreement with the model, the amplitude of the excited 
states are instead underestimated by the theoretical predictions. 
However, the analysis of the data is still ongoing. 
\\  
%
\section{Future Prospects}
For further studies of electroproduction of strangeness at JLab, a 
Letter of Intent (LoI 12-003) has been submitted to the Physics 
Advisory Committee (PAC39) and a new proposal is 
planned for the 12~GeV era at JLab. 
The future experiments will be performed in one experimental Hall only 
and the collaboration joining the previously independent Hall 
A and Hall C groups is studying the most favorable setup. \\
One scenario is the installation of the High-resolution Kaon 
Spectrometer (HKS) \cite{HashTam} in Hall A as $K^+$ arm. With 
the addition of two new septum magnets a setup similar to what used for 
E94-107 could be 
realized with the advantage of larger acceptance and smaller fraction of 
decayed $K^+$ in the spectrometer, which mean larger counting rates. 
With respect to E94-107, the kinematics should be adjusted in order to fit 
the new CEBAF specifications and the angular and momentum acceptance of HKS. 
The design of two new septum magnets and the definition of the optimal setup 
is under investigation and a proposal will be submitted accordingly, 
including also the study of the angular dependence of the elementary 
electroproduction of strangeness which would add data points in the 
unexplored region of Fig.~\ref{Fig1Label}. This study was 
already approved and tentatively scheduled (experiment E07-012) but it did 
not run in the 6-GeV era of JLab due to the incoming shutdown of the 
accelerator for the upgrade. \\
In the case of using a waterfall target, a simultaneous 
study of the angular dependence of the electroproduction of $^{16}_{\Lambda}N$ 
and elementary production could be performed. In addition, at very 
forward angles the ratio of the 
hypernuclear (calculated in DWIA) and elementary cross sections at the 
same 
kinematics should be almost independent of the used elementary 
amplitude, 
therefore the ratio contains direct information on the target and 
hypernuclear structure, production mechanisms and, possibly on the 
modification of the dynamics of the $p(e,e'K^+)$ process in the 
nuclear environment.\\
\begin{figure}[htb!]
\centering
\includegraphics[width=12.5cm]{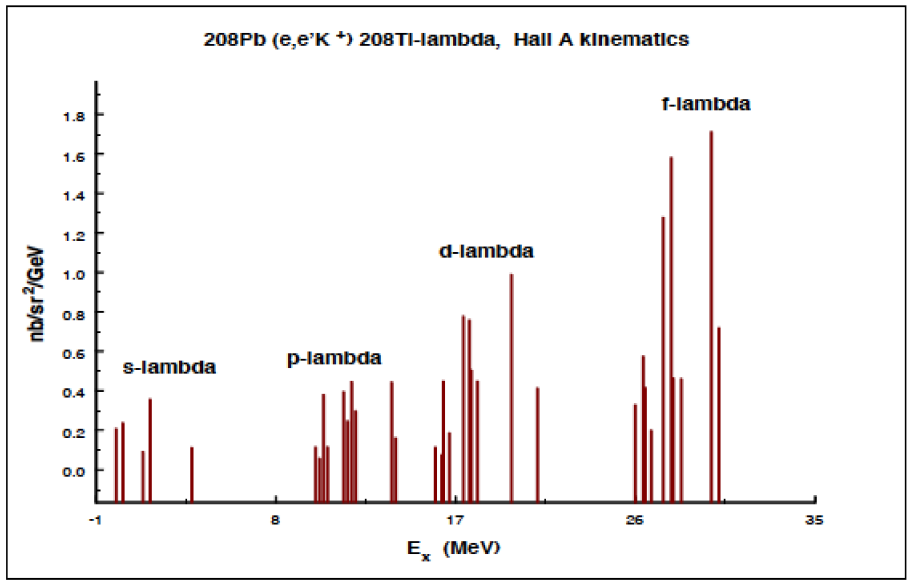}
\caption{\small \label{FigPb208}
Expected excitation energy spectrum for $^{208}_{\Lambda}Tl$, 
calculations performed by M. Sotona for the E94-107 kinematics.
}
\end{figure}
The future experiments might have sufficiently high 
counting rates to allow the study of medium-heavy or even as heavy 
hypernuclei as what could be produced on thin $^{208}Pb$ target. 
Very preliminary estimations of the counting rates of 
$^{208}_{\Lambda}Tl$ energy spectrum assuming a realistic setup in Hall 
A have been performed, based on the cross section reported in 
Fig.~\ref{FigPb208}.  
The calculations show that energy spectrum with statistics comparable 
to what is reported for example in \cite{Iodice} might be obtained in 
few weeks. 

\section{Conclusions}
The systematic study of hypernuclear spectroscopy by electroproduction 
of strangeness performed at Jefferson Laboratory has been very successful 
and has provided important 
elements for a better understanding of the baryon-baryon interactions 
and production mechanism in strangeness physics.
The planned study of angular dependence of the 
elementary electroproduction of strangeness in the very forward 
angular region is still not performed and 
it will be proposed again for the 12-GeV era. 
Furthermore, the possibility of investigating the structure 
of $^{208}_{\Lambda}Tl$ is under evaluation.

\section{Acknowledgments} We want to express a tribute to our colleague 
and friend Miloslav Sotona who gave a fundamental contribution to the 
history of the hypernuclear spectroscopy by electromagnetic probe.  
\\
Miloslav ``Sl\'avek" Sotona passed away on $6^{th}$ of April $2012$.

\end{document}